\documentclass[MSNbibl,nameyear,dvips]{arxstspdf}
\usepackage{flushend}
\usepackage{stfloats}
\volume{29}
\issue{2}
\pubyear{2014}
\firstpage{242}
\lastpage{246}
\doi{10.1214/14-STS471} % kopijuoti is 'New paper accepted'
\referstodoi{10.1214/13-STS457}% pagrindinio straipsnio DOI, kai straipsnis yra diskusija ar rejoinder'is



\begin{document}
\begin{frontmatter}

\vspace*{6pt}\title{Discussion of ``On the Birnbaum Argument for
the Strong Likelihood Principle''}%\thanksref{T1}
% kai straipsnis turi susijusiu diskusiju ir rejoinder'iu
\runtitle{Discussion}

\begin{aug}
\author[a]{\fnms{Michael}~\snm{Evans}\corref{}\ead[label=e1]{mevans@utstat.utoronto.ca}}
\runauthor{M. Evans}

\affiliation{University of Toronto}

\address[a]{Michael Evans is Professor, Department of Statistical Sciences,
 University of Toronto, 100 St. George St., Toronto, Ontario M5S 3G3, Canada \printead{e1}.}
\end{aug}

% ABSTRACT
\begin{abstract}
We discuss Birnbaum's result, its relevance to
statistical reasoning, Mayo's objections and the result in [\textit{Electron. J. Statist.} \textbf{7} (2013) 2645--2655] that
the proof of this result doesn't establish what is commonly believed.
\end{abstract}

% KEYWORDS
\end{frontmatter}

%s1 #&#
\section{Introduction}\label{sec1}

The result established in \citet{Bir62}, that if one accepts the frequentist
principles of sufficiency ($S$) and conditionality ($C$), then one must accept
the likelihood principle ($L$), has been an issue in the foundations of
statistics for 50 years. Many statisticians and philosophers of science accept
Birnbaum's theorem as a logical fact because the proof is simple and, if they
follow a pure likelihood or Bayesian prescription for inference, it doesn't
violate the way they think statistical analyses should be conducted. Many
frequentist statisticians reject the result basically because they don't like
the consequence that frequentist evaluations of statistical methodologies are irrelevant.

In the end, an acceptable theory of inference has to be based on sound logic
with no appeals to ex cathedra principles. Any principles used as
part of forming such a theory have to have strong justifications and produce
results that are free of paradoxes and contradictions. For example, the
principle of conditional probability, which says we replace $P(A)$ as the
measure of belief that event $A$ is true by $P(A \mid  C)$ after being told
that event $C$ has occurred, seems like a basic principle of inference that,
with careful application, is sound.

Does the likelihood principle carry the same weight in a theory of inference
as the principle of conditional probability? We don't think so and we will
later argue that a somewhat weakened version is really just a consequence of
the principle of conditional probability. Given that such principles can have
a significant influence on what we view as correct statistical reasoning, it
is important to examine the justifications for the likelihood principle, and
Birnbaum's theorem is commonly cited as such, to see if these are correct.

Another principle cited in Mayo's paper is the \textit{principle of
frequentism.} So what is the justification for this principle? Generally, this
seems to be based on the belief that it typically produces sensible
statistical methods, although, as we will subsequently discuss, the story seems
incomplete and unclear. If the principle of frequentism is correct, we need to
have a good argument for it and a much more complete development of the theory.

The relevance of frequentism to Mayo's paper lies in the author's position
that Birnbaum's argument is basically a violation of the principle. An
argument is provided for why the joint application of $S$ and $C$ used in
Birnbaum's proof constitute such a violation. I accept Mayo's reasoning. In
fact, I think it is somewhat similar to the argument put forward in \citet{EvaFraMon86N3} that the applications of $S$ and $C$ in the proof
are incorrect because $S$ discards as irrelevant precisely the information
used by $C$ to form the conditional model. So the justifications for $S$ and
$C$ contradict one another in the proof and this doesn't seem right. This
contradiction is avoided if one adopts the principle put forward in \citet{Dur70}, that we should restrict to ancillaries that are functions of a minimal
sufficient statistic, and then Birnbaum's proof fails.

As we will discuss in Section~\ref{sec3}, however, the issue in Birnbaum's argument is
not really with $S$ and $C$ together, but rather with $C$ itself and with what
is actually proved. A very broad hint that this is the case is provided in
\citet{EvaFraMon86N3} where, using the same style of argument as
Birnbaum, it is ``proved'' that accepting $C$ alone is equivalent to accepting
$L$. So Durbin's point doesn't save the day even if we accept it. Actually, I
don't think the arguments in Mayo's paper, or in \citet{EvaFraMon86N3}, completely dispense with Birnbaum's theorem either. They just
reinforce the unsettling feeling that something is wrong somewhere.

Section~\ref{sec3} contains an outline of \citet{r3} that, for me at least,
definitively settles the issue of what is wrong with Birnbaum's result and
does this mathematically. As will be apparent from Section~\ref{sec2}, however, it is
clear that I believe that a proper prior is a necessary part of formally
correct statistical reasoning. So why would a Bayesian want to invalidate
Birnbaum's result? This is because the result, as usually stated, is not
logically correct. Any valid theory has to have sound, logical foundations and
so we don't want any faulty reasoning being used to justify that theory. In
fact, Birnbaum's result even misleads, as we've heard it said that model
checking and checking for prior-data conflict violate the likelihood principle
and so should not be carried out. Both of these activities are a necessary
part of a statistical analysis. For this is how we deal, at least in part, with
the subjectivity inherent in a statistical analysis due to the choices made
by a statistician. This point, at least with respect to model checking, is
also made in Mayo's paper and I think it is an excellent one.

For proper Bayesians, a form of the likelihood principle is a consequence of
the principle of conditional probability, a far more important principle.
Applying the principle of conditional probability to the joint probability
model for the model parameter and data after observing the data, we have that
\textit{probability statements about the model parameter} depend on the
sampling model and data only through the likelihood (note the emphasis). Of
course, the likelihood map is minimal sufficient so there is nothing
surprising in this.

%s2 #&#
\section{Birnbaum and Evidence}\label{sec2}

There is an aspect of Birnbaum's work in this area that is particularly
noteworthy. This is his emphasis on trying to characterize statistical
evidence concerning the true value of the model parameter as expressed by the
function $Ev$. Consider the pairs $(M,x)$, where $M=\{f_{\theta}\dvtx \theta
\in\Theta\}$ is a set of probability distributions indexed by parameter
$\theta\in\Theta$ and $x$ is observed data coming from a distribution in $M$.
Then \citet{Bir62} writes $Ev(M_{1},x_{1})=Ev(M_{2},x_{2})$ to mean that the
evidence in $(M_{1},x_{1})$ is the same as the evidence in $(M_{2},x_{2})$
whenever certain conditions are satisfied$.$ We require here that $M_{1}$ and
$M_{2}$ have the same parameter space, but this can be weakened to
include models with parameter spaces that are bijectively equivalent.

The principles $S, C$ and $L$ are considered as possible partial
characterizations of statistical evidence. For example, if $(M_{1},x_{1})$ and
$(M_{2},x_{2})$ are related via $S$, then Birnbaum says that, for frequentist
statisticians, $Ev(M_{1},x_{1})=Ev(M_{2},x_{2})$ and similarly for $C$.
Birnbaum is careful to say that $Ev$ does not characterize what statistical
evidence is, it is a kind of ``equivalence relation'' (see Section~\ref{sec3}).

In essence Birnbaum brings us to the heart of the matter in statistical
inference. What is statistical evidence or, more appropriately, how do we
measure it? It seems collectively we talk about it, but we rarely get down to
details and really spell out how we are supposed to handle this concept. Perhaps
the closest to doing this is the pure likelihood theory, as discussed, for example, in \citet{r8}, but this is only a definition of relative evidence when comparing two
values of the full model parameter. For marginal parameters, this approach
uses the  profile likelihood as the only general way to compare the evidence
for different values and this is unsatisfactory from many points of view. For
example, a profile likelihood function is not generally a likelihood function.

For a frequentist theory of statistical inference, as opposed to a theory of
statistical decision, it seems essential that a general method for measuring
statistical evidence be provided that can be applied in any particular
problem. The $p$-value is often used as a frequentist measure of evidence
against a hypothesis, but, for a variety of reasons, it does not seem to be
appropriate. For example, we need a measure that can also provide evidence
\textit{for} something being true and not just evidence against, given that we
have assumed that the true distribution \textit{is} in $M$.

If we add a proper prior to the ingredients, then it seems we can come up with
sensible measures of evidence. For evidence, as expressed by observed data in
statistical problems, is what causes beliefs to change and so we can measure
evidence by measuring change in belief. For example, if we are interested in
the truth of the event $A$, and this has prior probability $P(A)>0$, then
after observing $C$, the principle of conditional probability leads to the
posterior probability $P(A \mid  C)$ as the appropriate expression of beliefs
about $A$. Accordingly, we measure evidence by the change in belief from
$P(A)$ to $P(A \mid  C)$. A simple \textit{principle of evidence} says that we
have evidence for the truth of $A$ when $P(A \mid  C)>P(A)$, evidence against
the truth of $A$ when $P(A \mid  C)<P(A)$ and no evidence one way or the other
when $P(A \mid  C)=P(A)$. This principle is common in discussions about evidence
in the philosophy of science and it seems obviously correct.

Of course, we also want to know how much evidence we have and this has led to
a variety of different measures based on $(P(A),P(A \mid  C))$. The Bayes factor
$\operatorname{BF}(A \mid  C)=P(A^{c})P(A \mid  C)P(A^{c})/P(A)P(A^{c} \mid  C)$ is one such
measure, as $\operatorname{BF}(A \mid  C)>1$ if and only if $P(A \mid  C)>P(A)$ and bigger values
mean more evidence in favor of $A$ being true. A central question associated
with this, and other measures of evidence, is how to calibrate its values, as
in when is $\operatorname{BF}(A \mid  C)$ big and when is it small. Actually, we prefer
measuring evidence via the relative belief ratio
$\operatorname{RB(}A \mid  C)=P(A \mid  C)/P(A)$, as the associated mathematics and the
calibration of its values are both simpler. The generalization to continuous
contexts is effected by taking limits and then both measures agree. A full
theory of inference, both estimation and hypothesis assessment, can be built
based on this measure of evidence together with a very natural calibration.
This is discussed in  \citet{r1}. Of course, many will not like
this because it involves proper priors, and so is subjective and supposedly
not scientific. Alternatively, some may complain that priors are somehow hard
to come up with.

In reality, all of statistics, excepting the data when it is properly
collected, is subjective. We choose models and we choose priors. What is
important is that any choice we make, as part of a statistical analysis, be
checkable against the objective data to ensure the choice at least make sense.
We check the model by asking whether or not the data is surprising for each
distribution in the model, and there are many well-known procedures for doing
this. Perhaps not so familiar is that we can also check a proper prior by
asking whether or not the true value is in a region of relatively low prior
probability. Procedures for doing this consistently are developed in \citet{r6} and \citet{r4}. In fact, there are even logical
approaches to modifying priors when prior-data conflict is found, as discussed
in \citet{r5}. Moreover, with a suitable definition of evidence,
we can measure a priori whether or not a prior is inducing bias into
a problem; see \citet{r1}. So subjectivity is not really the
issue. We do our best to assess and control its effects, and maybe that is
part of the role of statistics in science, but in the end it is an unavoidable
aspect of any statistical investigation.

It is undoubtedly true that it is possible to write down complicated models
for which it is extremely difficult, if not impossible, to prescribe an
elicitation procedure in an application that leads to a sensible choice of a
prior. But what does this say about our \textit{choice} of model? It seems
that we do not understand the effects of parameters in the model on the
measurements we are taking sufficiently well to develop such a procedure. That
is certainly possible, and perhaps even common, but it doesn't speak well for
the modeling process and it shouldn't be held up as a criticism of what should
be the gold standard for inference. An analogous situation arises with data
collection where we know the gold standard is random sampling from the
population(s) to which our inferences are to apply and, when we are interested
in relationships among variables, controlled allocation of the values of
predictors to sampled units. The fact that this is rarely, if ever, achieved
doesn't cause us to throw out the baby with the dirty bath water. Gold
standards serve as guides that we strive to attain and analyses that don't
just need to be suitably qualified.

Our main point in this section is that the problem of measuring statistical
evidence is the central issue in developing a theory of statistical
inference. It seems that Birnbaum realized this and was searching for a way to
accomplish this goal when he came upon what appeared to be a remarkable result.

%s3 #&#
\section{What's Wrong with Birnbaum's Result?}\label{sec3}

Perhaps everybody who has read the proof of Birnbaum's theorem is surprised at
its simplicity. In fact, this is one of the reasons it is so convincing, as
there does not appear to be a logical flaw in the proof. As Mayo has noted,
however, there are reasons to be doubtful of, if not even reject, the result
as being valid within the domain of any sensible theory of statistical
inference. Still suspicions linger, as the formulation seems so simple.

As we will now explain, the result proved is not really the result claimed. If
we want to treat Birnbaum's theorem and its proof as a piece of mathematics,
then we have to be precise about the ingredients going into it. It is the
imprecision in Birnbaum's formulation that leads to a faulty impression of
exactly what is proved. This is more carefully explained in \citet{r3}, but
we can give a broad outline here.

Suppose we have a set $D$. A relation $R$ on $D$ is any subset $R\subset
D\times D$. Meaningful relations express something and $(d_{1},d_{2})\in R$
means that $d_{1}$ and $d_{2}$ share some relevant property. Let $\mathcal{I}
$ denote the set of all model--data pairs $(M,x)$. So, for example, we can
consider $S$ as a relation on $\mathcal{I}$ by saying the pair $((M_{1}%
,x_{1}),(M_{2},x_{2}))\in S\subset\mathcal{I\times I}$ whenever $(M_{1}%
,x_{1})$ and $(M_{2},x_{2})$ have equivalent minimal sufficient statistics.
Similarly, $C$ and $L$ are relations on~$\mathcal{I}$.

An equivalence relation $R$ on $D$ is a relation that is reflexive: $(d,d)\in
R$ for all $d\in D$, symmetric: $(d_{1},d_{2})\in R$ implies $(d_{2},d_{1})\in
R$, and transitive: $(d_{1},\allowbreak d_{2}),(d_{2},d_{3})\in R$ implies $(d_{1}%
,d_{3})\in R$. It is reasonable to say that, whatever property is characterized
by relation $R$, when $R$ is an equivalence relation, then $(d_{1},d_{2})\in
R$ means that $d_{1}$ and $d_{2}$ possess the property to the same degree. It
is easy to prove that $S$ and $L$ are equivalence relations but $C$ and $S\cup
C$ are not equivalence relations; see \citet{r3}.

Associated with an arbitrary relation $R$ on $D$ is the smallest equivalence
relation on $D$ containing $R$, which we will denote by $\bar{R}$. Clearly,
$\bar{R}$ is the intersection of all equivalence relations containing $R$. But
$\bar{R}$ can also be characterized in another way that is key to Birnbaum's
proof.
\begin{Lemma*}\label{lem} If $R$ is a reflexive relation on $D$, then $\bar
{R}=\{(x,y)\dvtx \exists n,x_{1},\ldots,x_{n}\in D$ with $x=x_{1},y=x_{n}$ and
$(x_{i},x_{i+1})\in R$ or $(x_{i+1},x_{i})\in R\}$.
\end{Lemma*}
 Note that $S$ and $C$ are both reflexive and, thus, $S\cup C$ is reflexive.

In Birnbaum's proof, he starts with $((M_{1},x_{1}),\linebreak[4] (M_{2},x_{2}))\in L$,
namely, these pairs have proportional likelihoods. Birnbaum constructs
the mixture model (Birnbaumization) $M^{\ast}$ and then argues that we have that $((M_{1}%
,x_{1})$, $(M^{\ast},(1,x_{1})))\in C$, $((M^{\ast},(1,x_{1})),\allowbreak (M^{\ast}%
,(2,x_{2})))\in S$ and $((M^{\ast},(2,x_{2})),(M_{2},x_{2}))\in C$. Since
$C\subset S\cup C$ and $S\subset S\cup C$, by the \hyperref[lem]{Lemma}, this proves that
$L\subset\overline{S\cup C}$ and this is all that Birnbaum's argument
establishes. Since $S\cup C\subset L$ and $L$ is an equivalence relation, we
also have $L=\overline{S\cup C}$. As shown in \citet{r3}, it is also true that
$S\cup C$ is properly contained in $L$, so there is some content to the proof.
In prose, Birnbaum's proof establishes the following: if we accept $S$, and we
accept $C$, \textit{and} we accept all the equivalences generated by these
principles jointly, then we accept $L$. Certainly accepting $S$ and $C$ is not
equivalent to accepting $L$ since $S\cup C$ is a proper subset of $L$. We need
the additional hypothesis and there doesn't appear to be any good reason why
we should accept this as part of a theory of statistical inference. It is easy
to construct relations $R$ where $\bar{R}$ is meaningless. So we have to
justify the additional pairs we add to a relation when completing it to be an
equivalence relation.

It is interesting to note that the argument supposedly establishing the
equivalence of $C$ and $L$ in \citet{EvaFraMon86N3} also proceeds
in the same way using the method of the \hyperref[lem]{Lemma}. Since $C$ is properly contained
in $L$, this proof establishes that $\bar{C}=L$. So in fact, $S$ is irrelevant
in Birnbaum's proof. The problem with the principles $S$ and $C$, as partial
characterizations of statistical evidence, lies with $C$ and the fact that it
is not an equivalence relation. That $C$ is not an equivalence relation is
another way of expressing the well-known fact that, in general, a unique
maximal ancillary doesn't exist.

The result $\bar{C}=L$ does have some content. To be a valid characterization
of evidence in the context of $\mathcal{I}$, we will have to modify $C$ so
that it is an equivalence relation. The smallest equivalence relation
containing $C$ is $L$ and this is unappealing, at least to frequentists, as it
implies that repeated sampling properties are irrelevant for inference.
Another natural candidate for a resolution is the largest equivalence relation
contained in $C$ that is compatible with all the equivalence relations based
on maximal ancillaries. This is given by the equivalence relation based on the
laminal ancillary. From \citet{r2}, ancillary statistic $a$ is a
\textit{laminal ancillary} if it is a function of every maximal ancillary and
any other ancillary with this property is a function of $a$. The laminal
ancillary is essentially unique. It is unclear how appealing this resolution
would be to frequentists, but there don't seem to be any other natural candidates.

Many authors, including Mayo, refer to the weak conditionality principle which
restricts attention to ancillaries that are physically part of the
sampling. In such a case we would presumably write our models differently so
as to reflect the fact that this sampling occurred in stages. In other words,
the universe is different than $\mathcal{I}$, the one Birnbaum considered.
There doesn't seem to be anything controversial about such a principle and it
is well motivated by the two measuring instruments example and many others.

We don't believe, however, that the weak conditionality principle resolves the
problem with conditionality more generally. For example, how does weak
conditionality deal with situations like Example~2-2 in \citet{r7} and many
others like it? Conditioning on an ancillary seems absolutely essential if we
are to obtain sensible inferences in such examples, but there doesn't appear
to be any physical aspect of the sampling that corresponds to the relevant ancillary.

Many frequentist statisticians ignore conditionality, but, as noted in \citet{r7}, this is not logical. The theme in conditional inference is to
find the right hypothetical sequence of repeated samples to compare the
observed sample to. This takes us back to our question concerning the
principle of frequentism: why are we considering repeated samples anyway? A
successful frequentist theory of inference requires at least a resolution of
the problems with conditionality. The lack of such a resolution leads to
doubts as to the validity of the basic idea that underlies frequentism.

Issues concerning ancillaries are not irrelevant to Bayesians, as they have
uses in model checking and checking for prior-data conflict. Notice that the
principle of conditional probability does not imply that these activities need
refer to any kind of posterior probabilities and it is perfectly logical for
these to be based on prior probabilities. For example, model checking can be
based on the distribution of an ancillary or the conditional distribution of
the data given a minimal sufficient. Of course, $C$ is not relevant for proper
Bayesian probability statements about $\theta$, as the principle of conditional
probability implies that we condition on all of the data.

We acknowledge that it is possible that the problems with $C$ might be fixable
or even eliminated through a better understanding of what we are trying to
accomplish in statistical analyses---these aren't just problems in
mathematics. We can't resist noting, however, that the simple addition of a
proper prior to the ingredients does the job, at least for inference.

%s4 #&#
\section{Conclusions}\label{sec4}

Mayo's paper contains a number of insightful comments and, more generally, it
helps to focus attention on what is the most important question in statistics,
namely, what is the right way to formulate a statistical problem and carry out
a statistical analysis? To a certain extent, Birnbaum's result has been an
impediment in moving forward toward developing a theory of inference that has
a solid foundation. It is good to have such underbrush removed from the
discussion. We have to give great credit to Birnbaum, however, for his focus
on what is important in achieving this goal, namely, the measurement of
statistical evidence. That his theorem has lasted for so long is a testament
to the difficulties involved in this task.

In general, we need a strong foundation for a theory of statistical inference
rather than principles, often not clearly stated, that have only some vague,
intuitive appeal. The only way we can determine whether or not an instance of
statistical reasoning is correct lies within the context of a sound theory.
That two statistical analyses based on the same data and addressing the same
question can be deemed to be correct and yet come to different conclusions is
not a contradiction. Statistics tells us that we simply must collect more data
to resolve such differences. In our view, the role of statistics in science is
to explain how to reason correctly in statistical contexts. Without a strong
theory we can't do that.


%suskaldyti doi

% imsref loaded by jurgita.kaciuliene, 2014-05-28 11:19:11


\begin{thebibliography}{11}
% pybtex-1.08. Style name=ims, version=2.9, label_style=nameyear, sorting_style=complex, cfg=None, language=None.


%b1 ###
\bibitem[\protect\citeauthoryear{Baskurt and Evans}{2013}]{r1}
\begin{barticle}[mr]
\bauthor{\bsnm{Baskurt},~\bfnm{Zeynep}\binits{Z.}} \AND
\bauthor{\bsnm{Evans},~\bfnm{Michael}\binits{M.}}
(\byear{2013}).
\btitle{Hypothesis assessment and inequalities for {B}ayes factors and relative belief ratios}.
\bjournal{Bayesian Anal.}
\bvolume{8}
\bpages{569--590}.
\bid{doi={10.1214/13-BA824}, issn={1936-0975}, mr={3102226}}
\end{barticle}
\bptok{imsref}%
% NOT OUTPUTED:
%   issn = 1936-0975
%   url = http://dx.doi.org/10.1214/13-BA824
%   number = 3
%   fjournal = Bayesian Analysis
\endbibitem

%b2 ###
\bibitem[\protect\citeauthoryear{Basu}{1959}]{r2}
\begin{barticle}[mr]
\bauthor{\bsnm{Basu},~\bfnm{D.}\binits{D.}}
(\byear{1959}).
\btitle{The family of ancillary statistics}.
\bjournal{Sankhy\=a}
\bvolume{21}
\bpages{247--256}.
\bid{issn={0972-7671}, mr={0110115}}
\end{barticle}
\bptok{imsref}%
% NOT OUTPUTED:
%   issn = 0972-7671
%   fjournal = Sankhy\=a. The Indian Journal of Statistics
\endbibitem


%b3 ###
\bibitem[\protect\citeauthoryear{Birnbaum}{1962}]{Bir62}
\begin{barticle}[mr]
\bauthor{\bsnm{Birnbaum},~\bfnm{Allan}\binits{A.}}
(\byear{1962}).
\btitle{On the foundations of statistical inference}.
\bjournal{J. Amer. Statist. Assoc.}
\bvolume{57}
\bpages{296--326}.
\bid{mr={0138176}}
\end{barticle}
\bptok{imsref}%
% NOT OUTPUTED:
% issn = 0162-1459
% fjournal = Journal of the American Statistical Association
\endbibitem

%(\byear{1962}).
%% NOT OUTPUTED:
%% issn = 0162-1459
%% fjournal = Journal of the American Statistical Association


%b4 ###
\bibitem[\protect\citeauthoryear{Durbin}{1970}]{Dur70}
\begin{barticle}[auto:STB|2014/05/26|13:19:10]
\bauthor{\bsnm{Durbin},~\bfnm{J.}\binits{J.}}
(\byear{1970}).
\btitle{On Birnbaum's theorem on the relation between sufficiency, conditionality and likelihood}.
\bjournal{J. Amer. Statist. Assoc.}
\bvolume{65}
\bpages{395--398}.
\end{barticle}
\bptok{imsref}%
% NOT OUTPUTED:
% number = 329
\endbibitem


%b5 ###
\bibitem[\protect\citeauthoryear{Evans}{2013}]{r3}
\begin{barticle}[mr]
\bauthor{\bsnm{Evans},~\bfnm{Michael}\binits{M.}}
(\byear{2013}).
\btitle{What does the proof of {B}irnbaum's theorem prove?}
\bjournal{Electron. J. Stat.}
\bvolume{7}
\bpages{2645--2655}.
\bid{doi={10.1214/13-EJS857}, issn={1935-7524}, mr={3121626}}
\end{barticle}
\bptok{imsref}%
% NOT OUTPUTED:
%   issn = 1935-7524
%   url = http://dx.doi.org/10.1214/13-EJS857
%   fjournal = Electronic Journal of Statistics
\endbibitem

%b6 ###
\bibitem[\protect\citeauthoryear{Evans, Fraser and Monette}{1986}]{EvaFraMon86N3}
\begin{barticle}[mr]
\bauthor{\bsnm{Evans},~\bfnm{Michael~J.}\binits{M.~J.}},
\bauthor{\bsnm{Fraser},~\bfnm{Donald~A.~S.}\binits{D.~A.~S.}} \AND
\bauthor{\bsnm{Monette},~\bfnm{Georges}\binits{G.}}
(\byear{1986}).
\btitle{On principles and arguments to likelihood}.
\bjournal{Canad. J. Statist.}
\bvolume{14}
\bpages{181--199}.
\bid{doi={10.2307/3314794}, issn={0319-5724}, mr={0859631}}
\bptnote{check related}%
\end{barticle}
\bptok{imsref}%
% NOT OUTPUTED:
%   issn = 0319-5724
%   url = http://dx.doi.org/10.2307/3314794
%   number = 3
%   fjournal = The Canadian Journal of Statistics. La Revue Canadienne de Statistique
\endbibitem

%b7 ###
\bibitem[\protect\citeauthoryear{Evans and Jang}{2011a}]{r4}
\begin{barticle}[mr]
\bauthor{\bsnm{Evans},~\bfnm{Michael}\binits{M.}} \AND
\bauthor{\bsnm{Jang},~\bfnm{Gun~Ho}\binits{G.~H.}}
(\byear{2011a}).
\btitle{A limit result for the prior predictive applied to checking for prior-data conflict}.
\bjournal{Statist. Probab. Lett.}
\bvolume{81}
\bpages{1034--1038}.
\bid{doi={10.1016/j.spl.2011.02.025}, issn={0167-7152}, mr={2803740}}
\end{barticle}
\bptok{imsref}%
% NOT OUTPUTED:
%   issn = 0167-7152
%   url = http://dx.doi.org/10.1016/j.spl.2011.02.025
%   number = 8
%   coden = SPLTDC
%   fjournal = Statistics \& Probability Letters
\endbibitem\vadjust{\eject}

%b8 ###
\bibitem[\protect\citeauthoryear{Evans and Jang}{2011b}]{r5}
\begin{barticle}[mr]
\bauthor{\bsnm{Evans},~\bfnm{Michael}\binits{M.}} \AND
\bauthor{\bsnm{Jang},~\bfnm{Gun~Ho}\binits{G.~H.}}
(\byear{2011b}).
\btitle{Weak informativity and the information in one prior relative to another}.
\bjournal{Statist. Sci.}
\bvolume{26}
\bpages{423--439}.
\bid{doi={10.1214/11-STS357}, issn={0883-4237}, mr={2917964}}
\end{barticle}
\bptok{imsref}%
% NOT OUTPUTED:
%   issn = 0883-4237
%   url = http://dx.doi.org/10.1214/11-STS357
%   number = 3
%   fjournal = Statistical Science. A Review Journal of the Institute of Mathematical Statistics
\endbibitem

%b9 ###
\bibitem[\protect\citeauthoryear{Evans and Moshonov}{2006}]{r6}
\begin{barticle}[mr]
\bauthor{\bsnm{Evans},~\bfnm{Michael}\binits{M.}} \AND
\bauthor{\bsnm{Moshonov},~\bfnm{Hadas}\binits{H.}}
(\byear{2006}).
\btitle{Checking for prior-data conflict}.
\bjournal{Bayesian Anal.}
\bvolume{1}
\bpages{893--914 (electronic)}.
\bid{doi={10.1016/j.spl.2011.02.025}, issn={1931-6690}, mr={2282210}}
\end{barticle}
\bptok{imsref}%
% NOT OUTPUTED:
%   issn = 1931-6690
%   url = http://dx.doi.org/10.1016/j.spl.2011.02.025
%   number = 4
%   fjournal = Bayesian Analysis
\endbibitem

%b10 ###
\bibitem[\protect\citeauthoryear{Fraser}{2004}]{r7}
\begin{barticle}[mr]
\bauthor{\bsnm{Fraser},~\bfnm{D.~A.~S.}\binits{D.~A.~S.}}
(\byear{2004}).
\btitle{Ancillaries and conditional inference}.
\bjournal{Statist. Sci.}
\bvolume{19}
\bpages{333--369}.
\bid{doi={10.1214/088342304000000323}, issn={0883-4237}, mr={2140544}}
\bptnote{check related}%
\end{barticle}
\bptok{imsref}%
% NOT OUTPUTED:
%   issn = 0883-4237
%   url = http://dx.doi.org/10.1214/088342304000000323
%   number = 2
%   fjournal = Statistical Science. A Review Journal of the Institute of Mathematical Statistics
\endbibitem

%b11 ###
\bibitem[\protect\citeauthoryear{Royall}{1997}]{r8}
\begin{bbook}[mr]
\bauthor{\bsnm{Royall},~\bfnm{Richard~M.}\binits{R.~M.}}
(\byear{1997}).
\btitle{Statistical Evidence. A Likelihood Paradigm}.
\bseries{Monographs on Statistics and Applied Probability}
\bvolume{71}.
\bpublisher{Chapman \& Hall},
\blocation{London}.
\bid{mr={1629481}}
\end{bbook}
\bptok{imsref}%
% NOT OUTPUTED:
%   isbn = 0-412-04411-0
%   fpage = xvi+191
\endbibitem

\end{thebibliography}
\end{document}